\documentclass{article}
\usepackage{fortschritte}
\usepackage{amsmath,amsfonts,amssymb}
\usepackage{graphicx}
\usepackage{psfrag}
\usepackage{hyperref}
\def\beq{\begin{equation}}                     %
\def\eeq{\end{equation}}                       %
\def\bea{\begin{eqnarray}}                     %
\def\eea{\end{eqnarray}}                       %
\begin {document}
\def\titleline{
%
%
%
Gauge Instantons from Open Strings
}
\def\email_speaker{
{\tt
%
%
lerda@to.infn.it
}}
\def\authors{
%
%
%
%
M. Frau \1ad , A. Lerda \2ad{$^,$}\1ad\sp
}
\def\addresses{
%
%
%
%
\1ad
Dipartimento di Fisica Teorica, Universit\`a di Torino\\
and I.N.F.N., sezione di Torino\\
via P. Giuria 1, I-10125 Torino, Italy \\
\2ad
Dipartimento di Scienze e Tecnologie Avanzate\\
Universit\`a del Piemonte Orientale, I-15100 Alessandria, Italy\\
}
\def\abstracttext{
%
%
In this contribution we describe how to obtain
instanton effects in four dimensional
gauge theories by computing string
scattering amplitudes in D3/D$(-1)$ brane systems. In particular
we show that the disks with mixed boundary conditions, which are
typical of the D3/D$(-1)$ system, are the sources for the classical
instanton solution.
}
\large
\makefront
\section{Introduction}
In the last few years remarkable progress
in the study of (supersymmetric) field theories has been achieved
by embedding
them into string theory
and analyzing the infinite tension limit $\alpha'\to 0$.
In a perturbative framework, one typically considers string scattering
amplitudes on a Riemann surface $\Sigma$ of a given topology
\begin{equation}
{A}_N =  \phi_1\cdots \phi_N\,\int_{\Sigma}
\big\langle {\cal V}_{\phi_1}\cdots{\cal V} _{\phi_N}\big\rangle_{\Sigma}
\label{npoint}
\end{equation}
where the vertex operators $ {\cal V}_{\phi_i}$
describe the emission of the fields $\phi_i$
from the world-sheet, and $\langle\cdots\rangle_\Sigma$ denotes the
expectation value with respect to the vacuum represented by $\Sigma$.
For closed strings the simplest world-sheets are spheres, while
for open strings they are disks.
On these world-sheets there are no tadpoles:
in fact one has
\begin{equation}
\big\langle\,{\cal V}_{\phi_{\rm \,closed}}\,\big\rangle_{\rm sphere}
= 0
\label{vev0}
\end{equation}
for any closed string field $\phi_{\rm closed}$, and
\begin{equation}
\big\langle\,{\cal V}_{\phi_{\rm \,open}}\,\big\rangle_{\rm disk}
= 0
\label{vev}
\end{equation}
for any open string field $\phi_{\rm open}$. Thus, spheres and disks
are appropriate to describe the trivial vacua around which the ordinary
perturbation theory is performed, but are clearly inadequate
to describe classical non-perturbative backgrounds.

After the discovery of D-branes \cite{Polchinski:1995mt}
this perspective has drastically changed and nowadays several
non-perturbative features of field theory can be described in string theory.
The key observation is that, despite their non-perturbative nature,
the D-branes admit a perturbative description
in terms of closed strings whose left and right movers are suitably
identified. This amounts to insert a boundary on the world-sheet
so that the simplest topology for
closed strings in the presence of a D$p$ brane is that of disks
with $(p+1)$ longitudinal (or Neumann) and $(9-p)$ transverse
(or Dirichlet) boundary
conditions. As a consequence of the identifications between left and right
movers, on these disks there are closed-string tadpoles, {\it i.e.}
in general
\begin{equation}
\big\langle\,{\cal V}_{\phi_{\rm \,closed}}\,\big\rangle_{{\rm disk}_p}
\not= 0~~.
\label{vev1}
\end{equation}

A D$p$ brane can also be represented by a boundary state $|{\rm
D}p\rangle$, which is a non-perturbative state of the closed
string that inserts a boundary on the world-sheet and enforces on
it the appropriate boundary conditions
(for a review on the boundary state formalism, see for example
Ref.~\cite{DiVecchia:1999rh}). If we denote by $|\phi_{\rm
closed}\rangle$ the physical state associated to the vertex
operator ${\cal V}_{\phi_{\rm closed}}$, we have
\begin{equation}
\langle \phi_{\rm \,closed}|{\rm D}p\rangle
= \big\langle\,{\cal V}_{\phi_{\rm \,closed}}\,\big\rangle_{{\rm disk}_p}
~~.
\label{vev2}
\end{equation}
Thus, the boundary state, or equivalently its corresponding disk,
is a classical source for the various fields of the closed string spectrum.
In particular, it is a source for the massless fields (like
the graviton, or the Ramond-Ramond $(p+1)$-form potential)
which can acquire a non-trivial
profile and describe a classical background characterized by a
non-trivial geometry and a non-vanishing Ramond-Ramond charge.
The precise relation between such a background
and the boundary state has been established in
Ref.~\cite{DiVecchia:1997pr}, where it has been
shown that if one multiplies the massless tadpoles of
$|{\rm D}p\rangle$ by free propagators
and then takes the Fourier transform, one gets the leading terms
in the large distance expansion of the classical D$p$-brane solution.
These arguments prove that, in order to
describe closed strings in the non-perturbative
D-brane background it is necessary to
modify the boundary conditions of the string coordinates and, at
the lowest order, consider disks instead of spheres.

A natural question is whether this kind of approach can be
generalized to open strings, and in particular whether one can
describe in this way the instantons of four dimensional gauge
theories. A positive answer to this question has been given in
Ref.~\cite{Billo:2002}, and in this contribution we briefly review the
main results of that paper.
The crucial point is that the instantons of
(supersymmetric) gauge theories are
non-perturbative configurations which admit a perturbative
description within the realm of open string theory. Thus, they are
the analogue for open strings of what the
supergravity $p$-branes with Ramond-Ramond charge
are for closed strings.
In our analysis, a key role is played again by the
D-branes. For definiteness, let us consider a stack of $N$ D3
branes of Type IIB string theory which support on their
world-volume a four-dimensional
${\cal N}=4$ super Yang-Mills theory
with gauge group
${\mathrm{U}}(N)$~\footnote{or ${\mathrm{SU}}(N)$ if we disregard the
center of mass.}. Then, in order to describe
instantons of this gauge theory with topological charge $k$, one
has to introduce $k$ D$(-1)$ branes (D-instantons) and thus
consider a D3/D$(-1)$ brane system~\cite{Witten:1995im}.
In this system, besides the open strings
stretching between two D3 branes and representing the usual perturbative
gauge degrees of freedom, there are also open strings with at least one
end-point on the D-instantons. These strings describe non-dynamical
degrees of freedom which can be interpreted as
the moduli of the gauge (super)instantons in the
ADHM construction (for a review on this construction and its
realization using D-branes, see for instance Ref.~\cite{Dorey:2002ik}).

\begin{figure}[t]
\begin{center}
\includegraphics[width=0.10\textwidth]{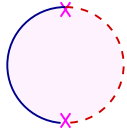}
\end{center}
\caption{The simplest mixed disk with two-boundary changing
operators indicated by the two crosses. The solid line represents
the D3 boundary while the dashed line represents the D$(-1)$
boundary.}
\label{fig:md0}
\end{figure}

In Ref.~\cite{Billo:2002} we have shown that this well-known
D-brane description is not only an efficient
book-keeping device to account for the multiplicities of
the various instanton moduli, but is
also a powerful tool to extract from string theory
a detailed information on the gauge instantons. The important
point is that in the D3/D$(-1)$ system the presence of
different boundary conditions for the open strings
implies the existence of disks whose boundary is
divided into different portions lying either on the D3 or on the
D$(-1)$ branes (see Fig. \ref{fig:md0}).
\noindent These disks, which we call
{\it mixed disks}, are characterized by the insertion of at least two
vertex operators associated to excitations of strings that stretch
between a D3 and a D$(-1)$ brane (or vice-versa), and
depend on the moduli that accompany
these mixed vertex operators. Due to the change in the
boundary conditions caused by the mixed operators, in general one
may expect that
\begin{equation}
\big\langle\,{\cal V}_{\phi_{\rm \,open}}\,\big\rangle_{\rm mixed~disk}
\not= 0~~.
\label{vev3}
\end{equation}
In Ref.~\cite{Billo:2002} we have shown that this expectation is indeed
correct. In particular the massless fields
of the ${\cal N}=4$ gauge vector
multiplet propagating on the D3 branes have non-trivial tadpoles
on mixed disks, and their Fourier transform (after including a propagator)
yields precisely the classical instanton solution of the ${\mathrm{SU(N)}}$
gauge theory in the singular gauge. In the next sections we briefly review
the main features of the D3/D($-1$) brane system, show how to obtain the
classical instanton profile from open string amplitudes on mixed disks,
and finally present our conclusions.

\section{The D3/D(-1) system}
As we mentioned above, the $k$ instanton sector of a four-dimensional
${\cal N}=4$ SYM theory with gauge group
$\mathrm{SU}(N)$ can be described by a bound state of $N$ D$3$ and
$k$ D$(-1)$ branes \cite{Witten:1995im}.
In the D3/D$(-1)$ system the string coordinates $X^{\cal
M}$ and $\psi^{\cal M}$ (${\cal
M}=1,\ldots,10$) obey different boundary conditions depending on
the type of boundary. Specifically, on the D$(-1)$ brane we have
Dirichlet boundary conditions in all directions, while on the D3
brane the longitudinal fields $X^\mu$ and $\psi^\mu$ ($\mu=1,2,3,4$)
satisfy Neumann boundary conditions, and the transverse fields
$X^a$ and $\psi^a$ ($a=5,\ldots,10$) obey Dirichlet boundary
conditions. These conditions, in turn, imply suitable reflection rules for
the spin fields,
which can be found in Ref.~\cite{Billo:2002}~\footnote{Here we recall that
the presence of the D3 branes breaks ${\mathrm{SO(10)}}$ into
${\mathrm{SO(4)}}\times {\mathrm{SO(6)}}$, so that the 10-dimensional spin
fields decompose into $S_\alpha S_A$ and $S^{\dot\alpha}S^A$
where $S_\alpha$ ($S^{\dot\alpha}$) are $\mathrm{SO}(4)$ Weyl spinors of
positive (negative) chirality, and $S^A$ ($S_A$) are SO(6) Weyl
spinors of positive (negative) chirality which transform in the
fundamental (anti-fundamental) representation of
$\mathrm{SU}(4)\sim \mathrm{SO}(6)$.}.
For our present purposes it is enough to specify the spectrum of
excitations of the open strings with at least one end-point on the
D-instantons. Let us first consider the strings that have
both ends on the D($-1$) branes: in the NS sector the physical
excitations are $a_\mu$ and $\chi^a$, whose corresponding vertex operators
are
\begin{equation}
\label{vertA}
V_a(z) \,=\,{a^\mu}\,\psi_{\mu}(z)
\,{\rm e}^{-\phi(z)} ~~~,
~~~
V_\chi(z)\,=\,{\chi^a}\,\psi_{a}(z)\,{\rm e}^{-\phi(z)}~~
\end{equation}
where $\phi$ is the chiral boson of the superghost bosonization.
In the R sector we find sixteen
fermionic moduli which are conventionally denoted by $M^{\alpha
A}$ and $\lambda_{\dot\alpha A}$, and correspond to the following
vertex operators
\begin{equation}
\label{vertM'} V_{M}(z)\,=\, M^{\alpha A}\,
S_{\alpha}(z)S_A(z)\,{\rm e}^{-\frac{1}{2}\phi(z)}~~~,
~~~
V_{\lambda}(z)\,=\,
{{\lambda_{\dot\alpha A}}}\,S^{\dot\alpha}(z)S^A(z)
\,{\rm e}^{-\frac{1}{2}\phi(z)}
~~.
\end{equation}
Let us now consider the open strings that start on a D3 and end on a
D($-1$) brane, or vice-versa. These strings
are characterized by the fact that four directions (the
longitudinal ones to the D3 brane) have mixed boundary conditions.
Thus, in the NS sector the four
fields $\psi^\mu$ have integer-mode expansions with
zero-modes that represent the $\mathrm{SO}(4)$ Clifford algebra and
the corresponding physical excitations are organized in two
bosonic Weyl spinors of $\mathrm{SO}(4)$. These are denoted by
$w_{\dot\alpha}$ and $\bar w_{\dot\alpha}$ respectively, and are
described by the following vertex operators
\begin{equation}
\label{vertexw} V_w(z) \,=\,{w}_{\dot\alpha}\, \Delta(z)\,
S^{\dot\alpha}(z) \,{\rm e}^{-\phi(z)}~~~,~~~
V_{\bar w}(z) \,=\,{\bar w}_{\dot\alpha}\, \bar\Delta(z)\,
S^{\dot\alpha}(z)\, {\rm e}^{-\phi(z)}~~.
\end{equation}
Here $\Delta(z)$ and $\bar\Delta(z)$ are the bosonic twist and
anti-twist fields with conformal dimension $1/4$, that change the
boundary conditions of the $X^\mu$ coordinates from Neumann to
Dirichlet and vice-versa by introducing a cut in the world-sheet.
In the R sector the fields
$\psi^\mu$ have, instead, half-integer mode expansions
so that there are fermionic zero-modes only in the six common
transverse directions. Thus, the physical excitations of this
sector form two fermionic Weyl spinors of ${\rm SO}(6)$ which are
denoted by $\mu^A$ and $\bar \mu^A$
respectively, and correspond to the following vertex operators
\begin{equation}
\label{vertexmu} V_\mu(z) \,=\,{\mu}^{A}\,
\Delta(z)\,S_{A}(z)\, {\rm e}^{-{1\over 2}\phi(z)}~~~,~~~
V_{\bar\mu}(z) \,=\,{{\bar \mu}}^{A}\,
\bar\Delta(z)\,S_{A}(z)\, {\rm e}^{-{1\over 2}\phi(z)}~~.
\end{equation}
A systematic analysis~\cite{Billo:2002} shows that, in the limit
$\alpha'\to 0$, the scattering amplitudes involving the above vertex
operators give rise to the following ``action''
\begin{equation}
\label{smoduli4}
\begin{aligned}
S&= {\rm tr}_k\Bigg\{
-\left[{a}^\mu, \chi^a\right]^2
+{\chi}_a\bar{w}_{\dot\alpha}{w}^{\dot\alpha}{\chi}^a
+\frac{{\rm i}}{2}\,\bar\Sigma^a_{AB}\,
\bar{\mu}^{A}{\mu}^{B}{\chi}_a-
\frac{{\rm i}}{4}\,\bar\Sigma^a_{AB}\,{M}^{\alpha A}\!
\left[{\chi}_a,{M}_{\alpha}^{~B}\right]
 \\
&~~+\,{\rm i}\Big(\bar{\mu}^{A}{w}_{\dot\alpha}+
\bar{w}_{\dot\alpha}{\mu}^{A} + \left[{M}^{\alpha
A},a_{\alpha\dot\alpha}\right]\Big){\lambda}^{\dot\alpha}_{~A}
-{\rm i}\Big(
w_{\dot\alpha}(\tau^c)^{\dot\alpha}_{~\dot\beta}\bar w^{\dot\beta}
+{\rm i}\,
\bar\eta_{\mu\nu}^c\left[{a}^\mu,{a}^\nu\right]\Big)D_c\Bigg\}
\end{aligned}
\end{equation}
where ${\rm tr}_k$ means trace over ${\mathrm U}(k)$,
$\bar\Sigma^a$ are the Dirac matrices of ${\mathrm{SO}}(6)$,
$\bar\eta^c$ ($c=1,2,3$) are the anti-self dual 't Hooft symbols,
and $\tau^c$ are the Pauli matrices. In (\ref{smoduli4})
there appear also three
auxiliary fields $D_c$ whose string representation
is provided by the following vertex operators~\cite{Billo:2002}
\begin{equation}
\label{vertaux}
V_D(z) \,=\, \frac{1}{2}
D_c\,\bar\eta_{\mu\nu}^c\,\psi^\nu(z) \psi^\mu(z)~~.
\end{equation}
As is well known~\cite{Dorey:2002ik}, by simply taking ${\rm e}^{-S}$
one obtains the measure on the instanton moduli space,
while by varying $S$ with respect to $D_c$ and
${\lambda}^{\dot\alpha}_{~A}$ one easily derives the bosonic and fermionic
ADHM constraints.

\section{The instanton solution from mixed disks}
\begin{figure}[t]
\begin{center}
\psfrag{mu}{ $A_\mu^c$}
\psfrag{p}{\small $p$}
\psfrag{w}{\small $\bar w$}
\psfrag{wb}{\small $w$}
\includegraphics[width=0.21\textwidth]{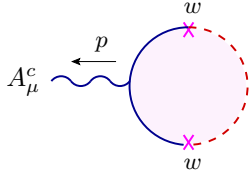}
\end{center}
\caption{The mixed disk that describes the emission of a gauge
vector field $A_\mu^c$ with momentum $p$ represented by the
outgoing wavy line.}\label{fig:md2}
\end{figure}

We now show that the mixed disks of the D3/D($-1)$ brane system
(see Fig. \ref{fig:md0}) are the source for the instanton
background of the super Yang-Mills theory, and that the classical
instanton profile can be obtained from open string amplitudes. For
simplicity we will discuss only the case of instanton number $k=1$
in a $\mathrm{SU}(2)$ gauge theory, but no substantial changes
occur in our analysis if one considers higher values of $k$ and
other gauge groups (see Ref.~\cite{Billo:2002} for these
extensions). Let us then consider the emission of the
$\mathrm{SU}(2)$ gauge vector field $A_\mu^c$ from a mixed disk.
The simplest diagram which can contribute to this process contains
two boundary changing operators $V_{\bar w}$ and $V_{w}$ and no
other moduli, and is shown in Fig. \ref{fig:md2}.
\noindent
The amplitude associated to this diagram is
\begin{equation}
\label{dia1}
\left\langle {\cal V}_{A^c_\mu(-p)}\right\rangle_{\rm mixed~disk}
 \,\equiv\,
\left\langle V_{\bar w}\,{\cal V}_{A^c_\mu(-p)}\,V_{w}
\right\rangle \,=\,  {\cal A}^c_\mu(p ;{\bar w, w})
\end{equation}
where the gluon vertex operator ${\cal V}_{A^c_\mu(-p)}$
is with {\it outgoing} momentum
$p$ and {without} polarization, so that the amplitude carries the
structure of an emitted gauge vector field. The evaluation of the
amplitude (\ref{dia1}) is quite straightforward and the result
is~\cite{Billo:2002}
\begin{equation}
\label{ampl}
{\cal A}^c_\mu(p ;{\bar w, w}) = {\rm i}\rho^2\, p^\nu\,\bar\eta^c_{\nu\mu}
\,{\rm e}^{-{\rm i} p\cdot x_0}
\end{equation}
where we have defined $\bar w^{\dot\alpha} w_{\dot\alpha}=2\rho^2$
and denoted by $x_0$ the location of the D-instanton inside the world-volume
of the D3 branes. By taking the Fourier transform of (\ref{ampl}), after
inserting a free propagator~\cite{DiVecchia:1997pr}, we obtain
\begin{equation}
{\cal A}_\mu^c(x)
\,=\,\int\frac{d^4p}{(2\pi)^2}\,\frac{1}{p^2}\,{\cal A}^c_\mu(p ;{\bar w, w})
\,{\rm e}^{{\rm i} p\cdot x} \,=\,
\frac{2\rho^2\,\bar\eta^c_{\mu\nu}\,(x-x_0)^\nu}{(x-x_0)^4}~~.
\label{solution}
\end{equation}
In (\ref{solution}) we recognize the leading term in the
large distance expansion ({\it i.e.} $|x-x_0|>\!>\rho$) of the
$\mathrm{SU}(2)$ instanton
with center $x_0$ and size $\rho$ in the
\emph{singular gauge}, namely
\begin{equation}
{\cal A}_\mu^c(x) = \frac{2\rho^2 \,\bar\eta^c_{\mu\nu}(x -
x_0)^\nu}{ (x - x_0)^2 \Big[(x-x_0)^2 + \rho^2\Big]}\simeq
\frac{2\rho^2 \,\bar\eta^c_{\mu\nu}\,(x - x_0)^\nu}{ (x -
x_0)^4}\left(1 - {\rho^2\over (x-x_0)^2} + \ldots\right)~.
\label{solution1}
\end{equation}
This result shows that a mixed disk, like that of Fig.
\ref{fig:md2}, is the source for the classical gauge instanton.
Note that the amplitude (\ref{dia1}) is a 3-point function from
the point of view of the two dimensional conformal field theory on
the string world sheet, but is a 1-point function from the point
of view of the four-dimensional gauge theory on the D3 branes. In
fact, the two boundary changing operators $V_{\bar w}$ and $V_{w}$
that appear in (\ref{dia1}) just describe non-dynamical parameters
on which the background depends. Furthermore, the fact that the
gluon field (\ref{solution}) is in the singular gauge is not
surprising, because in our set-up the gauge instanton is produced
by a D$(-1)$ brane which is a point-like object inside the D3
brane world-volume. Thus it is natural that the gauge connection
exhibits a singularity at the location $x_0$ of the D-instanton.

An obvious question at this point is whether also the subleading
terms in the large distance expansion (\ref{solution1}) have a
direct interpretation in string theory. Since such terms contain
higher powers of $\rho^2\sim \bar w^{\dot\alpha} w_{\dot\alpha}$,
one expects that they are associated to mixed disks with more
insertions of boundary changing operators. This expectation has
been explicitly confirmed in Ref.~\cite{Billo:2002}, so that one
can conclude that mixed disks with the emission of a gauge vector
field do reproduce the complete $k=1$ $\mathrm{SU}(2)$ instanton
solution. As mentioned above, this result can be extended to other
instanton numbers (and other gauge groups) in a straightforward
way.

\section{Conclusions}
In this contribution we have considered, for simplicity, only
the emission of gluons from mixed disks; however, this
approach can be easily generalized
to study the emission of the complete vector supermultiplet
and recover in this way the full superinstanton
solutions of the ${\cal N}=4$ super Yang-Mills theory~\cite{Billo:2002}.
Furthermore, this analysis can be
extended without difficulties also to D3/D$(-1)$
brane systems in orbifold backgrounds that
reduce the supersymmetry to ${\cal N}=2$ or ${\cal N}=1$, as well as
to brane systems in the
presence of constant NS-NS or R-R background fields.

In conclusion we have have shown that the mixed disks are the sources for
the gauge (super)instantons, and thus, they are the appropriate
world-sheets one has to consider in order to compute instanton
contributions to correlation functions within string theory.
We believe that this result helps to clarify the analysis and the
prescriptions presented for example in Ref.~\cite{Green:1997tv}
and also provides the conceptual bridge necessary to relate the
D-instanton techniques of string theory to the
instanton calculus of field theory.
\\ \\
{\bf Acknowledgements:} We would like to deeply thank
our coauthors: M. Bill\`o, F. Fucito, A. Liccardo and I. Pesando.
This work is partially supported by the European Commission RTN programme
HPRN-CT-2000-00131 and by MIUR under contract 2003-023852.

\end{document}